\documentclass[12pt,preprint]{aastex}

\newcommand{\pg} {PG 1234$+$482}
\newcommand{\logg} {\log g}
\newcommand{\Te} {T_{\rm eff}}

\newcommand{\hbeta} {H$\beta$}
\newcommand{\hgamma} {H$\gamma$}
\newcommand{\hdelta} {H$\delta$}

\shortauthors{Pereira et al.}
\shorttitle{Spectroscopic Variations in GD 323}
\begin{document}

\title{DISCOVERY OF SPECTROSCOPIC VARIATIONS IN THE DAB WHITE DWARF GD 323}

\author{C. Pereira, P. Bergeron, and F. Wesemael}
\affil{D\'epartement de Physique, Universit\'e de Montr\'eal, C.P.~6128, 
Succ.~Centre-Ville, Montr\'eal, Qu\'ebec, Canada, H3C 3J7.}
\email{pereira@astro.umontreal.ca, bergeron@astro.umontreal.ca, 
wesemael@astro.umontreal.ca}

\begin{abstract}

We report the discovery of spectroscopic variations in GD 323, the
prototypical DAB white dwarf. Simultaneous optical spectroscopic
observations over five consecutive nights of GD 323 and of
\pg, a non-variable comparison DA white dwarf of similar
brightness, are used to reveal quasi-periodic variations in both the
hydrogen and helium absorption lines over a timescale of hours. The
amplitude of the variation of the equivalent width of \hbeta\ is
$\sim30$ \%.  Moreover, the strength of the hydrogen lines is shown to
vary in opposite phase from that of He~\textsc{i} $\lambda$4471. These results
suggest that the model currently thought to be the most viable to
account for the simultaneous presence of hydrogen and helium lines in GD 323, namely
a static stratified atmosphere, may need to be reexamined. Instead, a
model with an inhomogeneous surface composition, resulting perhaps from
the dilution of a thin hydrogen atmosphere with the underlying helium
convection zone, may be a better representation of GD 323.  The
observed variation timescale of $\sim3.5$ hours is consistent with the
slow rotation rate of white dwarf stars.

\end{abstract}


\section{INTRODUCTION}

The DAB stars represent a class of white dwarfs with hybrid spectra, as
weak neutral helium lines are superposed onto the classical
hydrogen-line spectrum of DA stars \citep [see, e.g.,][]{atlas}. The
prototype of this class, GD 323 (WD $1302+597$; $V=14.52$), was
discovered independently by \citet{owk84} and \citet{liebertetal84}.
The latter made the first attempts at modeling the unusual hybrid
spectrum of GD 323, and their results have since been updated and
considerably upgraded by \citet{kls94} on the basis of a set of
high-quality optical spectra of that star. \citet{kls94} were led to
conclude that a stratified atmospheric structure, consisting of a thin
(${\rm log}\,M_{\rm H}/M_{\star} = -16.8$) hydrogen layer floating on
top of a helium envelope, remained the ``most promising'' explanation
of the bulk of the properties of GD 323.

Since this initial discovery, several additional stars
spectroscopically classified as DAB stars have been unearthed:
G104$-$27 \citep{hkw90}, HS 0209$+$0832 \citep{jhek93}, MCT 0128$-$3846
and MCT 0453$-$2933 \citep{wesemaeletal94}, PG~1115$+$166
\citep{burleighetal01, bl02} and, most recently, PG~1603$+$432
\citep{vdc04}. The sample of DAB stars is quite inhomogeneous, however,
as several distinct mechanisms have been called on to account for
individual objects: thus, GD 323 could be a stratified object, while
it has been argued in the discovery papers cited above that MCT
0128$-$3846, MCT 0453$-$2933, and PG~1115$+$166 are unresolved DA$+$DB
binaries. This has now been confirmed for PG~1115$+$166
\citep{maxtedetal02} and MCT 0453$-$2933 \citep{napiwotzkietal05}.
Furthermore, HS 0209$+$0832, while initially thought to be
characterized by a homogeneous atmosphere \citep{hnle97}, now appears
to be undergoing accretion \citep{wolffetal00}. The last object added
to the list, PG 1603$+$432, seems characterized by a homogeneous
composition.

Among those stars, the prototype GD 323 is perhaps the most thoroughly
studied object. In their detailed analysis, \citet{kls94} investigate
four distinct models: homogeneous and stratified composition,
unresolved binary and spotted star. Despite this variety of
options, they conclude that none of these models provides a completely
satisfactory fit to the data, but that the stratified model is the
most acceptable match. Given this ambiguous outcome, further
investigations of these models appear in order.

An important ingredient in defining the correct model for GD 323 could
be the question of variability. While the white light photometry of
\citet{rw83} excludes photometric variations at the 0.25\% level in
the range 10--1200 s, it was suggested by \citet{liebertetal84} that
the object be monitored for spectroscopic variations. This need became
more acute with the report by \citet{kidderetal92} that the weak He I
lines, initially observed by \citet{hkw90} in data at high S/N on the
DAB star G104-27, were not seen in later spectra of comparable quality.
A similar situation is encountered in HS 0209$+$0832, in which
\citet{hnle97} report changes in the strength of the neutral and
ionized helium lines over a one-year period. In GD 323, earlier spectra
obtained by \citet{liebertetal84} and by J.~L.~Greenstein --- initially
displayed by \citet{koester91a, koester91b} and \citet{atlas} ---
unfortunately cannot provide any firm conclusion about the variability
of that object, as these were obtained with different instrumental
setups and spectroscopic resolutions.

To investigate whether GD 323 is a spectroscopic variable,
\citet{kls94} obtained a series of nine high signal-to-noise ratio,
$\sim7$ \AA\ resolution spectra of that object, all secured with an
essentially identical instrumental setup. Six of these spectra were
obtained within a single night, while the other three had been secured,
respectively 15 days, 9 months, and one year earlier. On the basis of
these spectra, the best and most homogeneous set secured at that time
for a DAB star, \citet{kls94} concluded that the available data were
compatible with the assumption that the star showed no variability.
However, it was our feeling at the time that, while it was clear that
there were no large-amplitude changes in the line spectrum of GD 323, the
signal-to-noise ratio of the individual spectra was good enough to
permit a much more detailed analysis of the variability of GD 323 than
had been carried out by \citet{kls94}. This was undertaken
independently by \citet{wesemaeletal95}, but their analysis could not
quite meet the burden of proof associated with an investigation of this
type: while their results suggested the presence of small spectroscopic
variations in the \citet{kls94} data, their analysis also
showed that a well-planned observing strategy and a rigorous data
reduction process would both be required in order to build a stronger
case for the presence, or absence, of low-level spectroscopic
variations in GD 323.

Recently, the opportunity arose to revisit the issue of spectroscopic
variability of GD 323, and we succeeded in securing new data for that
object that sheds some light on this 20 year-old problem. This paper
summarizes the results of this revamped investigation.

\section{OBSERVATIONS}

\subsection{Data Acquisition}

As discussed above, the study of \citet{kls94} reveals that the
spectroscopic variations in GD 323, if any, are quite small.  Hence a
careful observing strategy had to be developed to ensure that any
variability detected in a series of time-resolved spectroscopic
observations is intrinsic to the star and not the result of changes in
the atmospheric transparency or of an artifice introduced in the data
reduction. The best way to assess the variability of GD 323 is to
secure in parallel spectroscopic observations of a constant comparison
star. Ideally, one would like to have both GD 323 and the comparison
star on the slit in order to obtain simultaneous time-resolved
spectroscopic observations. Unfortunately, no such star could be found
in the vicinity of GD 323, and we had to rely on a different strategy.

We thus selected a comparison star based on the following set of
criteria:  firstly, we required a star with no known variations. We
also required an object whose magnitude was similar to that of GD 323
($V=14.52$). Furthermore, we demanded that our comparison object have
few spectral lines and a spectral energy distribution comparable to
that of GD 323. Finally, we wanted an object that was as close as
possible to our target object in the sky. Examination of the Catalog of
Spectroscopically Identified White Dwarfs of \citet{mccook99} revealed
that the DA star \pg\ was the most judicious choice.  Past
observations of \pg\ have shown no evidence for variability, it
has a visual magnitude of $V=14.42$, and with an effective temperature
of $\Te=55,040$~K \citep[][hereafter LBH]{liebert05}, its spectrum consists of a
well-defined continuum with only weak hydrogen lines. Finally,
\pg\ and GD 323 are separated by approximately 15 degrees on
the sky.

Since we did not know {\it a priori} over what timescales variations
might occur, we carried out spectroscopic observations over five
consecutive nights, from 2004 February 10 to 14. Both GD 323 and
\pg\ were monitored for over 5 hours on the first night, and
for about an hour on the remaining nights. These optical spectra were
secured using the Steward Observatory 2.3 m reflector telescope
equipped with the Boller \& Chivens spectrograph and a Loral CCD
detector. A 4.5 arcsec slit and a 600 line mm$^{-1}$ grating in first
order provided a spectral coverage of 3200-5300~\AA\ at a resolution of
$\sim$ 6~\AA\ FWHM. Each 600 s spectroscopic observation of GD 323 was
immediately followed by a 600 s exposure of \pg. The average
signal-to-noise ratio per pixel and per exposure is $\sim 85$ for GD
323 and $\sim100$ for \pg. In total, 27 spectra of GD 323 and
25 spectra of \pg\ were secured, the vast majority of them
under excellent observing conditions.

\subsection{Data Reduction}

The optical spectra were extracted and wavelength-calibrated using the
Image Reduction and Analysis Facility (IRAF) standard package. A first
cut at the flux calibration was obtained with IRAF using the various
flux standards secured during the observing nights. However, in the
course of our analysis, we realized that we could take advantage of our
multiple spectroscopic observations of \pg, and use this star
instead as a flux standard, at least in a relative sense. This method
is analogous to that used for taking high-speed photometric
observations of variable stars, where a constant comparison star is
used to correct the light curve of the target star. The detailed
procedure used in the reduction of our spectroscopic observations thus
proceeded as follows.

The first step was to combine our 25 individual spectra of \pg,
flux calibrated with IRAF as described above. This combined spectrum,
characterized by a signal-to-noise ratio of $\sim340$, was then fitted
with a grid of synthetic white dwarf spectra, from which we derived
$\Te=54,200$~K and $\logg=7.77$. This effective temperature is within
1.5 \% from that derived by LBH on the basis of a lower
signal-to-noise ratio spectrum. A synthetic spectrum calculated with
these atmospheric parameters, which matches extremely well the combined
observed spectrum, now serves as a noiseless template for the rest of
the analysis. Flux calibration functions were then obtained by simply
dividing this template by the individual unfluxed spectra of
\pg\ and smoothing the result with a 9-point filter. The
application of these smoothed flux calibration functions to the
unfluxed spectra of \pg\ ensures that each resulting spectrum
matches the synthetic template perfectly. The fluxed spectra for
\pg\ are shown in the right panel of Figure \ref{fg:f1} for
the first night, and in the right panel of Figure \ref{fg:f2} for the
remaining nights.

The next step was to use these derived functions to flux-calibrate the
spectroscopic observations of GD 323. Specifically, we calibrate each
spectrum of GD 323 with the function derived from the observation of
\pg\ that immediately follows. The resulting flux-calibrated
spectra of GD 323 are displayed in the left panels of Figures
\ref{fg:f1} and \ref{fg:f2}. There are two main advantages to our
approach: first we obtain flux calibrated spectra of our target star
that are as independent as possible of the flux calibration provided by
IRAF. Here, the IRAF package is used only to generate the average
spectrum of \pg\ used to define the noiseless synthetic
template of that star. Once this template is defined, the uncoupling
from IRAF in the procedure that generates calibrated spectra of GD 323
is complete (IRAF spectra are used to generate our error estimate in
\S\ 3).  This is a substantial advantage, given that the flux
calibration is recognized to be the most delicate step of the reduction
procedure. In addition, because of the nearly simultaneous observations
of GD 323 and \pg, our method allows us to monitor small
temporal changes in the atmospheric transparency.

As will be discussed further below, our conclusions about the
variability of GD 323 do not hinge on the flux calibration procedure
developed here, although the quality of the results has certainly
benefited from this approach. One should also note that we did not
worry about calibrating our spectra in terms of absolute fluxes, as we
are mainly interested here in relative flux variations.

\section{A SEARCH FOR SPECTROSCOPIC VARIATIONS IN GD 323}

\subsection{Single-Night Variations in Equivalent Widths}

With our calibrated spectra in hand, we search for spectroscopic
variations by measuring the equivalent widths of the most important
hydrogen and helium absorption lines. The most delicate step in this
procedure is to define the continuum in order to normalize the
individual line profiles. A simple average over a few pixels on each
side of the line is not accurate enough and can introduce undesirable
uncertainties in the equivalent width measurements. We rely instead on
a procedure similar to that outlined in LBH to define the
location of the continuum in the far wings of the hydrogen Balmer
lines.

We first construct a very high signal-to-noise template spectrum by
combining our 27 spectroscopic observations of GD 323. This template
spectrum displayed at the top of Figure \ref{fg:f2} is then forced to
match each single observation of GD 323 by multiplying the template by
a polynomial of fifth degree in $\lambda$. The result of this procedure
for a typical spectrum of GD 323 is illustrated in Figure \ref{fg:f3}.
We define the continuum level over a given interval on the basis of the
monochromatic flux in the template spectrum at the boundaries of the
interval. We choose the following intervals over which the equivalent
widths are to be measured: 4800-4925~\AA, 4275-4425~\AA, 4065-4150~\AA,
corresponding to \hbeta, \hgamma, and \hdelta, respectively, as well as
4430-4510~\AA\ and 4680-4750~\AA\ for the He~\textsc{i} $\lambda$4471
and He~\textsc{i} $\lambda$4713 features, respectively. We concentrate
on these features since they are the strongest lines in the spectrum;
other lines are visible, but are either weak or blended and can only be
measured with difficulty. Furthermore, the wavelength intervals are
chosen to include as much of the line wings as possible, while avoiding
blending. The intervals over which the equivalent widths are measured
are reproduced in Figure \ref{fg:f3}.

The equivalent width measurements as a function of time for the three
hydrogen and the two neutral helium absorption lines in GD 323 are
displayed in Figure \ref{fg:f4} for the 14 spectra secured within the
single night of 2004 February 10.  These results show clearly that the
hydrogen and He~\textsc{i} $\lambda$4471 line strengths of GD 323 vary with time,
and that these variations appear to be quasi-periodic. This was by no
means expected, as we could very well have detected random variations,
or no variations at all. Moreover, the observed variations occur on a
relatively short timescale, with the feature strengths going from a
maximum to a minimum over the course of a few hours only. Even more
surprising, the variations of the hydrogen lines and $\lambda$4471 appear
to be out of phase: as the hydrogen lines (especially \hbeta\ and
\hgamma) get stronger, the helium
line weakens and vice versa. The case for He~\textsc{i} $\lambda$4713
is admittedly much weaker and we cannot, at this stage, account
completely for its behavior: while the line is reasonably strong and
unaffected by blending, the variations it displays appear more muted
than those observed in $\lambda$4471, and their phasing with the
hydrogen lines much less obvious.

\subsection{Error Analysis}

A preliminary estimate of the errors in the equivalent width
measurements for the three hydrogen lines in GD 323 has been obtained
from a consideration of the spectrum of the comparison star
\pg. We have measured the equivalent widths for each line on
the 13 individual fluxed spectra of that star secured on 2004 February
10. The rms error on the equivalent widths reflects the S/N ratio of
the individual spectra and the reliability of the procedure used to
set the continuum level (\S\ 2). They amount to 0.03 \AA\ for
H$\beta$, 0.08 \AA\ for H$\gamma$, and 0.03 \AA\ for H$\delta$. These
errors can be considered as lower limits, as the use of smoothed flux
calibration functions applied to the raw individual spectra of \pg\
guarantees a perfect match of the continuum flux in individual spectra
with the template of that star.  When the same smoothed flux
calibration functions are applied to the raw individual spectra of GD
323, the match to the average, or template, spectrum of GD 323 is not
as good, since residual calibration errors remain. It is not possible
to evaluate these calibration errors, however, as they cannot be
disentangled from the intrinsic variations of GD 323.

To estimate these residual calibration errors, we have opted to go
back to the 13 individual IRAF-reduced spectra of \pg\ acquired on
2004 February 10. Because these spectra were simply reduced with the
standard IRAF package, the rms deviations over a single night are
deemed to include contributions from the S/N ratio of the individual
spectra, from the continuum setting and, more importantly, from any
residual calibration error associated with IRAF. While our standard
reduction is decoupled from IRAF (\S\ 2.2), it is our feeling that the
rms errors generated in this way are more representative of the true
uncertainties associated with our procedure. Their values are: 0.27
\AA\ for H$\beta$, and 0.16 \AA\ for H$\gamma$ and H$\delta$.

These IRAF-based error bars are included in Figure \ref{fg:f4}.  For
the neutral helium lines $\lambda$4471 and $\lambda$4713, both absent
in the spectrum of \pg, we also measure the rms errors for
H$\epsilon$ (0.20 \AA), the Balmer line that is closest in average
equivalent width to the two weaker He~\textsc{i} lines.

\subsection{Single-Night Variations in Calibrated Spectra}

An alternative way of illustrating these variations is to divide the
template spectrum of GD 323 (shown in Fig.~3) by each observed
spectrum. Were GD 323 a constant star, this procedure should yield a
straight line with random noise. The results for 3 spectra obtained on
2004 February 10 are displayed in Figure \ref{fg:f5} at carefully
selected phases. The bottom spectrum corresponds to a phase where the
hydrogen line equivalent widths go through a minimum in Figure
\ref{fg:f4}, while the top one corresponds to a maximum, and the
middle spectrum is for an intermediate value; the trend is in the
opposite direction for the helium lines. Also shown are the wavelength
intervals used to measure the equivalent widths. Our results indicate
that the flux variations never exceed 5 \% throughout the spectrum. A
close examination reveals, however, that in the case of the 1043
spectrum, there is less flux in the hydrogen features and more flux in
the He~\textsc{i} features, while the opposite is true for the 1226
spectrum, even when taking into account the noise level. By
integrating these small variations over the entire line profiles, we
reduce the effect of the noise, and increase the significance of these
variations.  Indeed, even a 1\% flux variation over a 100 \AA\ wide
\hbeta\ line profile yields a 1 \AA\ equivalent width variation, which
is roughly the size of the variations observed in Figure \ref{fg:f4}.

\subsection{Single-Night Variations in IRAF-calibrated and Uncalibrated Spectra}

To reinforce these conclusions, we have carried out in parallel an
analysis of the spectroscopic observations of GD 323 and \pg, but this
time on data that were flux calibrated with IRAF. An additional
analysis was also carried out directly with the unfluxed spectra. In
both cases, the quasi-periodic variations in the hydrogen and helium
features uncovered in \S\ 3.1 could also be detected, although the
quality of the results was significantly reduced. The fact that we
were able to achieve similar results via several methods makes it
unlikely that the variations uncovered in this paper are due to some
artifice in the data acquisition or reduction procedures.

\subsection{Variations over longer timescales}

The equivalent width measurements for all five successive nights --- a
total of 27 observations --- are shown in Figure \ref{fg:f6} for the
aforementioned features. The variations observed on the first night
persist on the following nights, although only a segment of the
quasi-sinusoidal variations is observed. Remarkably, on a given night,
all hydrogen lines follow a similar pattern of variations, while the
helium lines also share a common pattern, different from that of the
hydrogen lines. Our attempts to fit a single periodic function through
all the data points have failed, indicating that the variations are
only quasi-periodic.

\section{DISCUSSION}

The spectroscopic variations uncovered in GD 323 occur on
timescales of the order of 3.5 h. Were these variations associated with
the rotation of the underlying white dwarf ($R=1.4\times
10^{-2}\,R_\odot$), they would correspond to an equatorial velocity
$v=5\,{\rm km\,s^{-1}}$. This number is quite in line with the idea
that most white dwarfs of both DA and DB types are very slow rotators
\citep{hnr97, koesteretal98, dwb02, karletal04}. This suggests that the
association of the 3.5 h timescale with stellar rotation is
reasonable.

It is also possible to revisit the issue of secular spectroscopic
variations on timescales of years in the spectrum of GD 323. The first
digital optical spectra of GD 323 were obtained over 20 years ago, but
the different instrumental setups and resolutions used always hampered
a direct comparison with spectra secured more recently
\citep{liebertetal84, kls94}. However, the setup for the observations
of \citet{kls94} and ours is quite similar and this provides a better
basis for the comparison of high S/N ratio spectra secured 11 years
apart. A comparison of the average spectrum of \citet{kls94} for the
night of 1993 March 31 (average of 6 spectra) with our so-called
template spectrum (average of 27 spectra) is shown in Figure
\ref{fg:f7}. Although the reduction procedures differ, the claim can
certainly be made that there are no large-amplitude secular variations
in the spectrum of GD 323 over that baseline. However, small
variations, most notably in He~\textsc{i} $\lambda$4026 and
He~\textsc{i} $\lambda$4471, may be present.

The detection of spectroscopic variations on a 3.5 h timescale in GD
323 forces us to revise our ideas about this unusual object. As
discussed in the Introduction, the thorough analysis of \citet{kls94}
suggested that none of the options they considered provided a
completely satisfactory fit to the data, but the stratified model
provided the most acceptable match to GD 323. It seems difficult to
understand now how the spectroscopic variations uncovered here could
be accommodated by a static, layered model or, for that matter, by a
static homogeneous hydrogen-rich atmosphere containing traces of
helium.

Our findings suggest instead that an inhomogeneous surface abundance
distribution in a slowly rotating star should be favored. Models with
inhomogeneous surface abundances were discussed by
\citet{beauchampetal93}, who considered several belt and cap
geometries, and by \citet {kls94}, who considered a spot model.  Can
such models reproduce the basic characteristics of GD 323 ? Let us
consider the conclusions of \citet{kls94} first: they model the spot
geometry by adding appropriately weighted fluxes from model atmospheres
of DA and DB stars and restrict, on physical grounds, the difference in
effective temperatures of these models to less than 5000 K. The
conclusion they derive from the analysis of spotted models is a
familiar one \citep [see, e.g.,][]{liebertetal84}; namely that it is difficult
to find a single model (characterized here by the effective temperature
of the hydrogen-rich and helium-rich spots and by the relative surface
they cover) that matches both the hydrogen and helium line strength
and the slope of the energy distribution of GD 323.

In an earlier analysis, \citet{beauchampetal93} had used more
sophisticated models that considered specific geometries (equatorial
belts and polar caps) and included a self-consistent treatment of the
limb darkening effects. Sample optical spectra of models with a pure
helium equatorial band and pure hydrogen polar caps from that
investigation are reproduced here in Figure \ref{fg:f8} and contrasted
with our average spectrum of GD 323. These results illustrate how the
strength of the hydrogen and helium lines, and in particular their
relative strength, depends on both the extent of the helium-rich belt
and on the inclination of the stellar symmetry axis with respect to
the line of sight (given here by the angle $\theta_{\rm axis}$).  For
a given half-width of the pure helium equatorial band of
$\Delta\alpha=45^\circ$, the hydrogen lines are stronger and the
helium lines weaker when the star is seen pole-on ($\theta_{\rm
axis}=0^\circ$) instead of equator-on ($\theta_{\rm
axis}=90^\circ$). Similarly, for a given inclination axis of
$\theta_{\rm axis}=0^\circ$ (pole-on), the helium lines become
stronger and the hydrogen lines weaker when the half-width of the pure
helium equatorial band is increased from $\Delta\alpha=45^\circ$ to
$60^\circ$.  In the present context, variations of the hydrogen and
helium line strengths in opposite phase could be reproduced if the
rotation axis and the symmetry axis are misaligned.  While no optimal
fit is provided here or in the investigation of
\citet{beauchampetal93}, the results of Beauchamp et al.~appear more
encouraging than those of \citet{kls94}. In particular, they
encountered no major difficulty in matching the energy distribution
with a model that gave an acceptable, although admittedly far from
perfect, fit to the average optical spectrum of GD 323.

The modeling of surface inhomogeneities in terms of spots, or caps and
belts, can perhaps lead one to believe that the origin of these
surface features must be directly related to the presence of a
magnetic field at the stellar surface. In that picture, GD 323 would
be a low-field analog of Feige 7, a DAB white dwarf with a 35 MG
dipolar field \citep{achilleosetal92}. In GD 323, the null circular
polarization measurement of \citet{abl81} yields a $3\sigma$ upper
limit of 3 MG on the longitudinal field component at the stellar
surface. However, the idea that surface inhomogeneities are related to
other processes, such as the convective dilution of a thin overlying
hydrogen envelope, should also be entertained. Indeed, it has been
pointed out a while back \citep{lfw87} that GD 323 is slightly cooler
than the red edge of the DB gap, that peculiar region between 45,000 K
and 30,000 K that appears to be devoid of helium-atmosphere
degenerates.  Models of the spectral evolution of white dwarfs \citep
[e.g.,][]{fw87} attempt to account for this gap by calling on the
dilution of a thin hydrogen layer at the surface of a 30,000 K DA star
with the more massive helium envelope, dilution that would lead to a
DA$\to$DB transition near that effective temperature.  If GD 323 were
understood in those terms, an idea that we favor, its surface might be
characterized by the presence of horizontal abundance
gradients. Unfortunately, it is not possible to be more specific since
a complete picture of the mixing process still eludes us at this
stage. The geometries considered in the exploratory investigations of
\citet{beauchampetal93} and \citet{kls94} appear reasonable starting
points for a renewed and more thorough study of that possibility.

We thank the director and staff of the Steward Observatory for the use
of their facilities, and A.~Beauchamp for his essential contribution to
the calculation of models with surface inhomogeneities. This work
was supported in part by the NSERC Canada and by the Fund FQRNT
(Qu\'ebec).

\clearpage

\figcaption[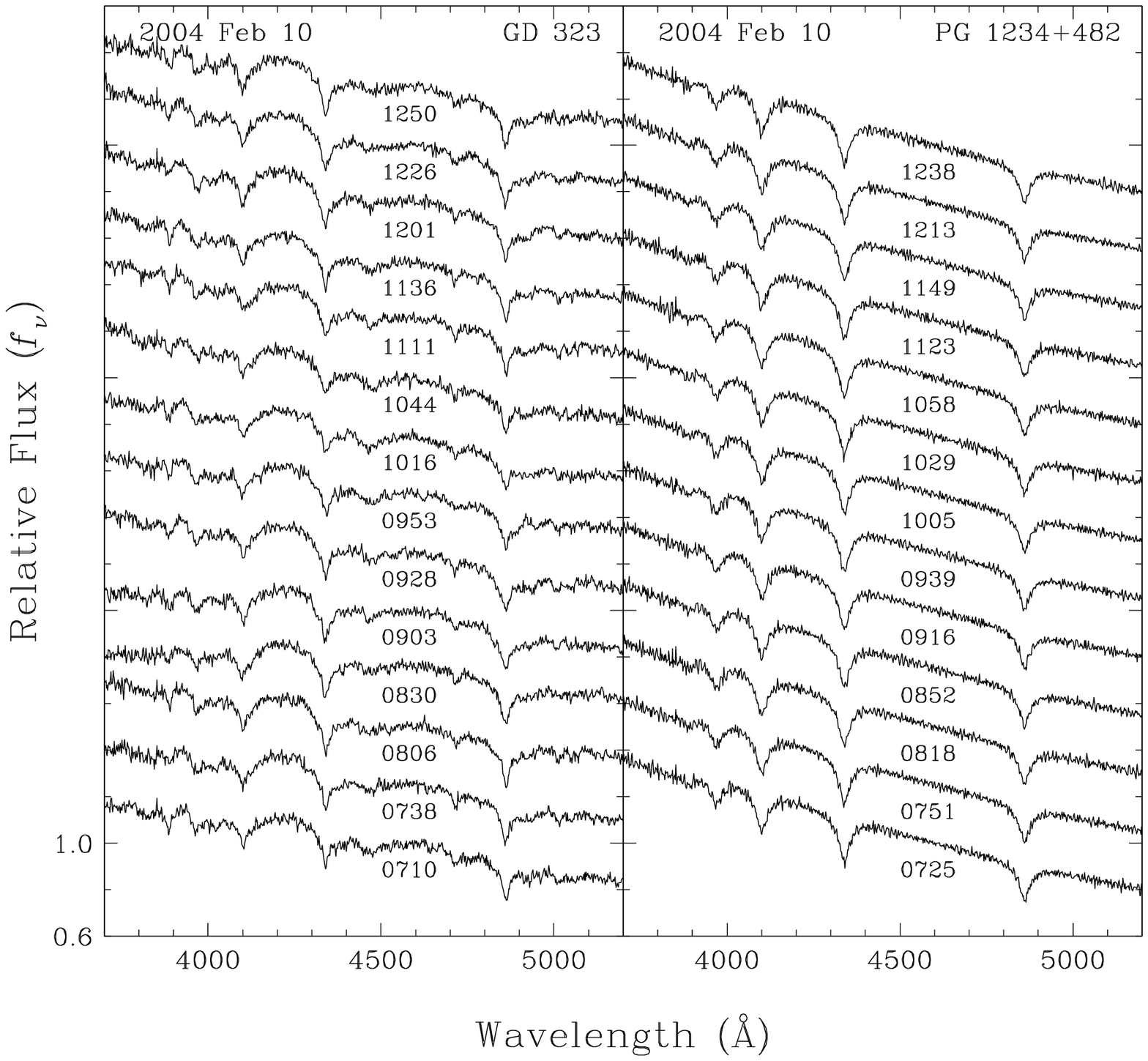] {Optical spectra of GD 323 ({\it left panel}) 
and \pg\ ({\it right panel}) taken on 2004 February 10. The
spectra appear in chronological order from the bottom up and are
labeled with the corresponding universal time at which the integration
was started. The spectra have been flux calibrated, normalized to
unity at 4600~\AA\ and offset from each other by a factor of 0.25 for
clarity.\label{fg:f1}}

\figcaption[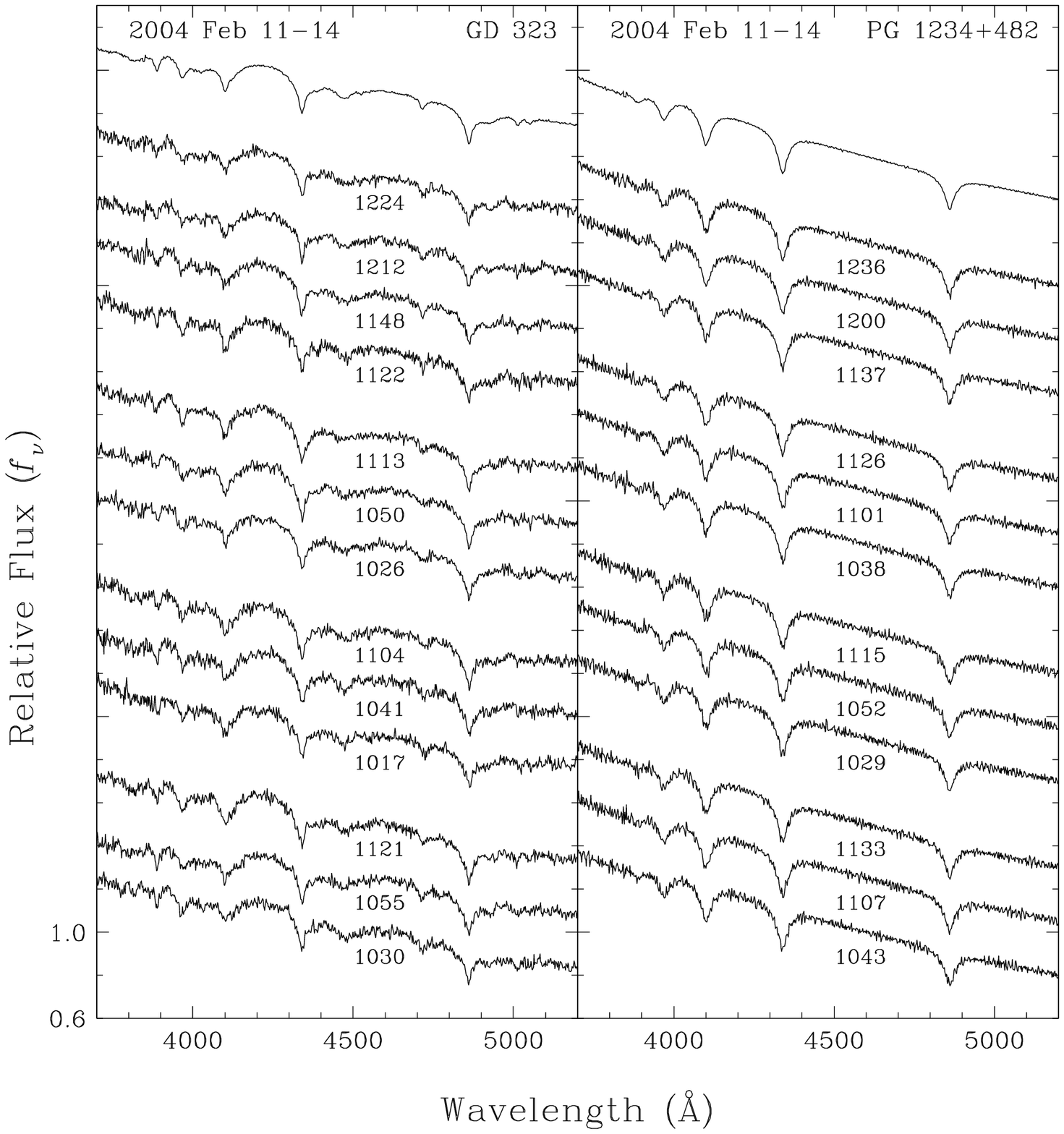] {Optical spectra of GD 323 ({\it left panel}) 
and \pg\ ({\it right panel}) taken on 2004 February 11-14. The
spectra appear in chronological order from the bottom up and are
labeled with the corresponding universal time at which the integration
was started. The spectra have been flux calibrated, normalized to
unity at 4600~\AA\ and offset from each other by a factor of 0.25
within the same night and by 0.3 between consecutive nights.  The
combined spectrum of GD 323 from all nights is also displayed at the
top of the left panel; for comparison, the combined spectrum of
\pg\ is displayed at the top of the right
panel.\label{fg:f2}}

\figcaption[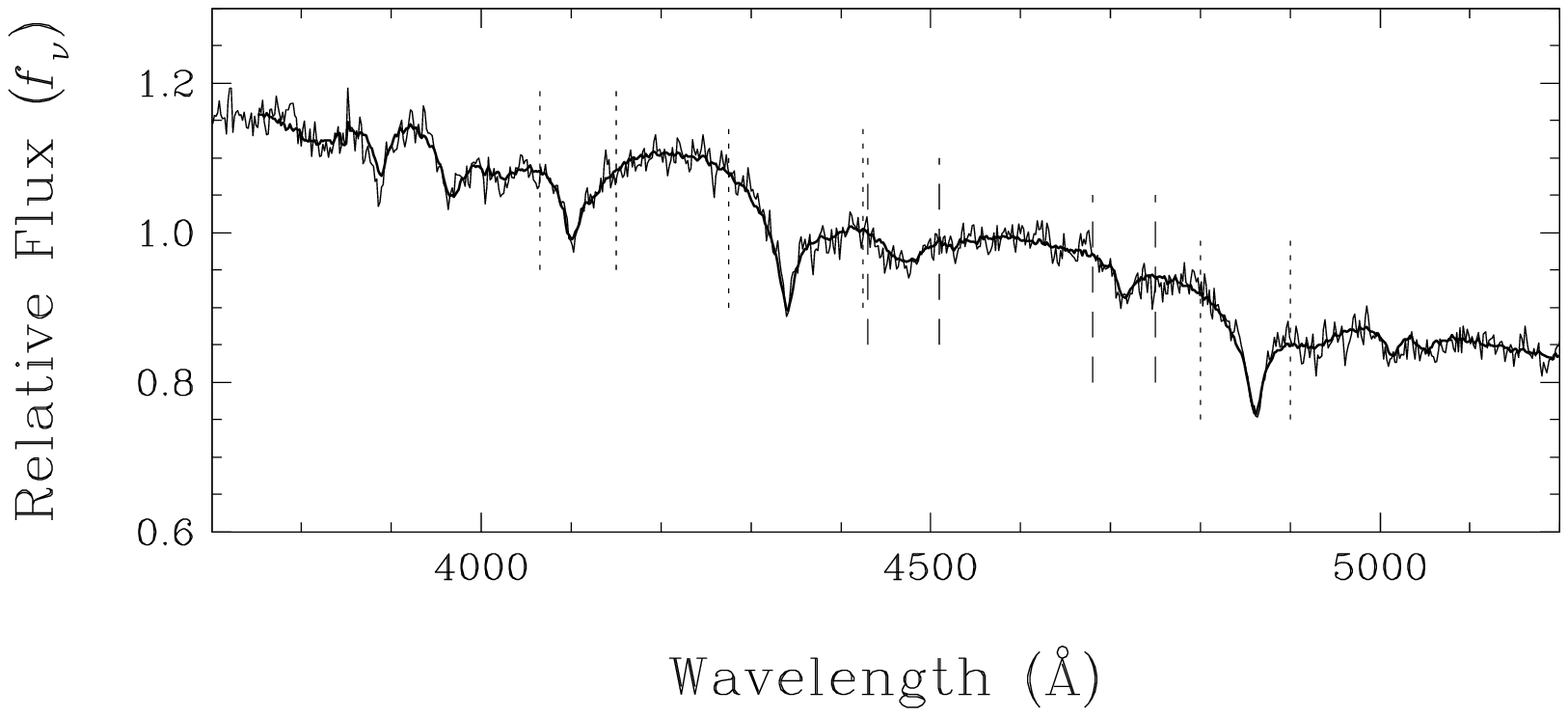]{Template spectrum (average of 27 spectra) 
superimposed on a typical spectrum of GD 323. Here, the template
continuum has been adjusted to that of the observed spectrum using the
procedure described in the text. This is done for each individual
spectrum. The wavelength range used to measure the equivalent widths
is shown for the hydrogen ({\it dotted lines}) and the He~\textsc{i}
({\it dashed lines}) absorption lines.\label{fg:f3}}

\figcaption[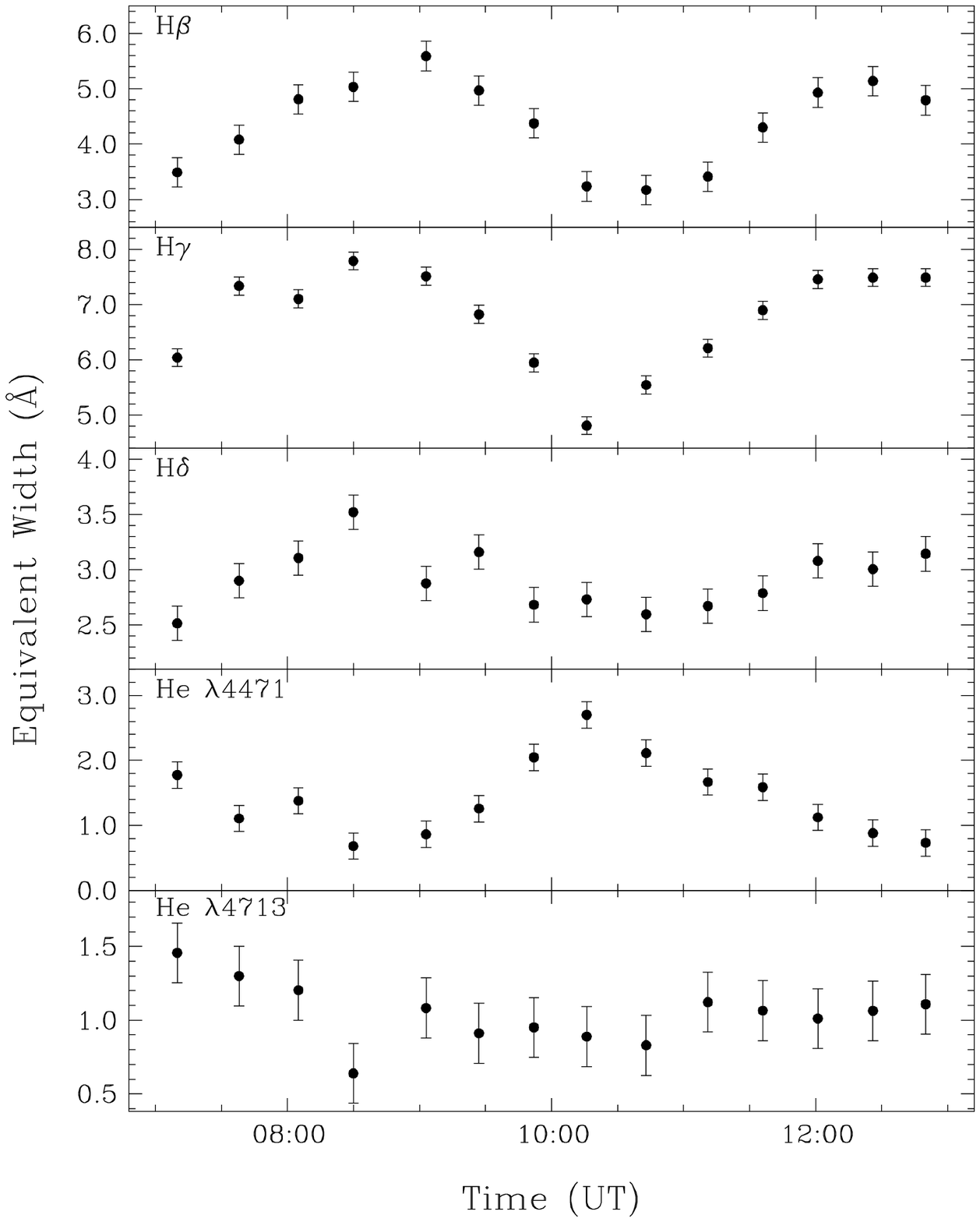]{Equivalent widths as a function of time (hours UT) 
for the spectra of GD 323 taken on 2004 February 10. The panels
correspond to \hbeta, \hgamma, \hdelta, He~\textsc{i} $\lambda$4471
and He~\textsc{i} $\lambda$4713 from top to bottom. The error bars are
estimated from the equivalent width measurements of \pg, which
showed no significant variations and had a similar signal-to-noise
ratio. These values are then used as error bars for GD
323. \label{fg:f4}}

\figcaption[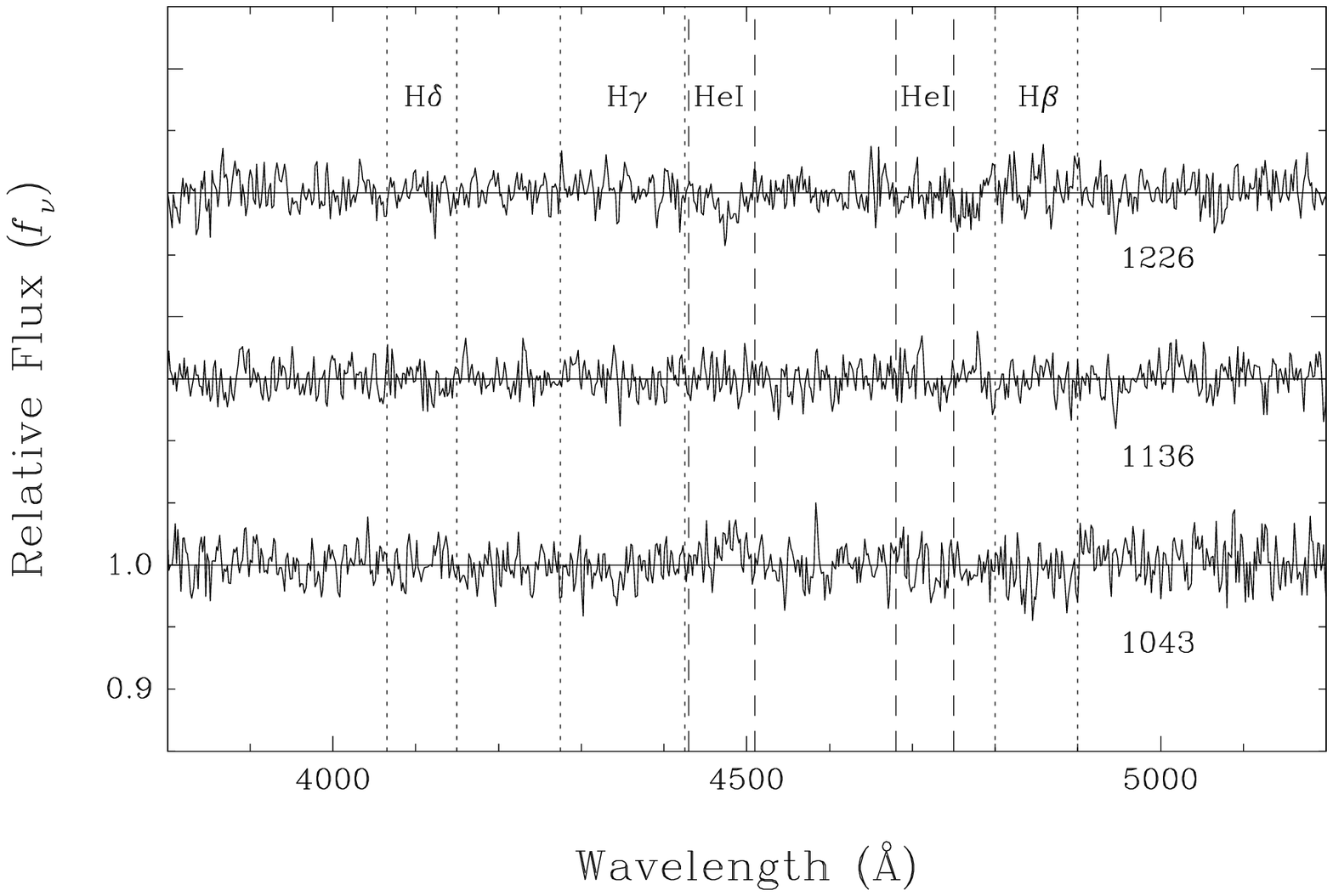]{Results obtained by dividing the fitted template 
spectrum of GD 323 (see, e.g., Fig.~\ref{fg:f3}) by the observed
spectra. The spectra are all taken on 2004 February 10, and they are
labeled with the universal time of their acquisition. The spectra
selected correspond to various phases where the hydrogen line
strengths go from a minimum to a maximum, from bottom to top. The
wavelength range used for the equivalent width measurements are
indicated for the five strong features:
\hbeta-\hdelta~({\it dotted lines}), He~\textsc{i} $\lambda$4471 and
He~\textsc{i} $\lambda$4713 ({\it dashed lines}). \label{fg:f5}}

\figcaption[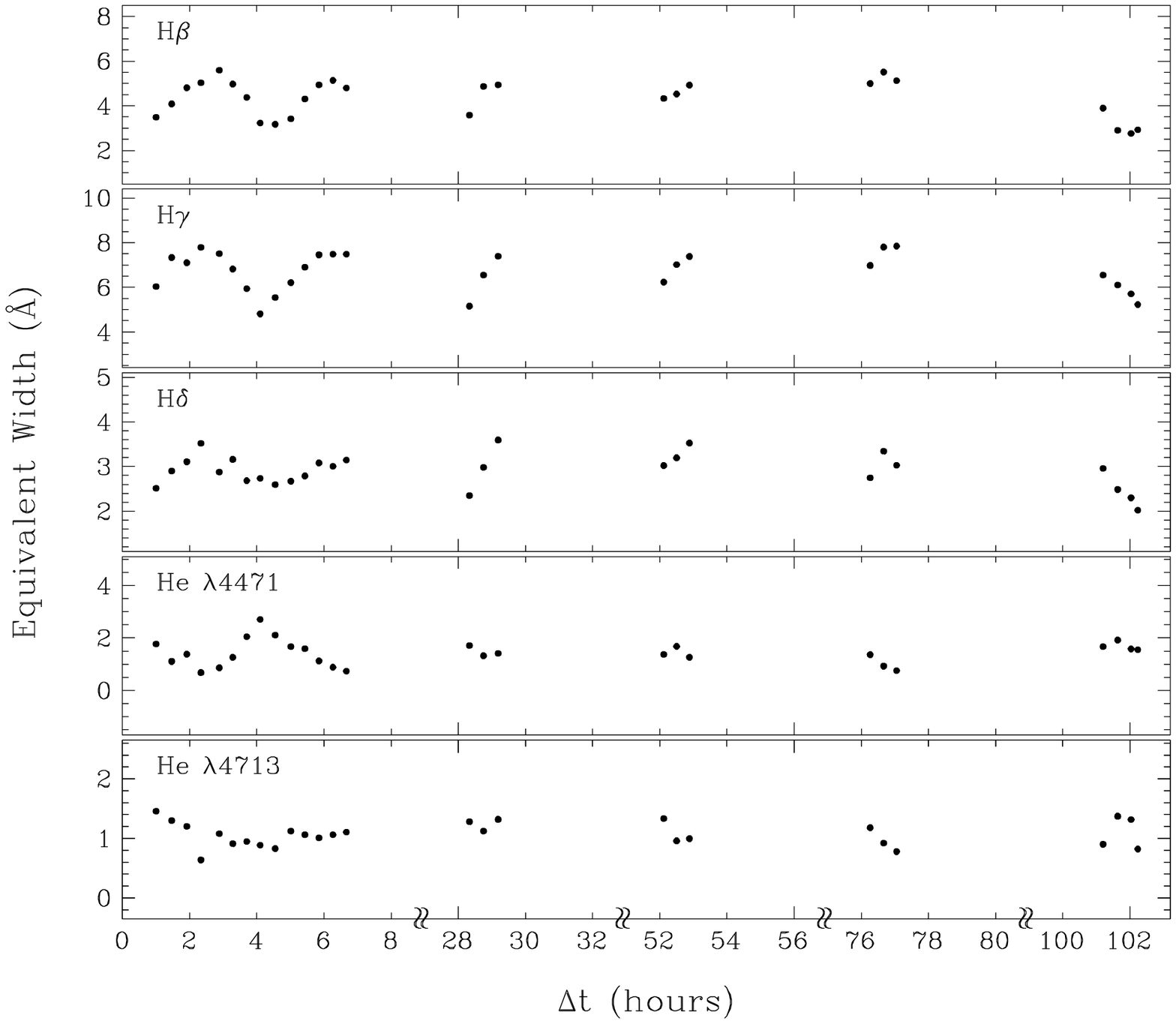]{Equivalent widths as a function of relative time (in hours) 
for the spectra of GD 323 taken over five nights, namely from 2004
February 10-14. The first observation is offset from $t=0$ by one
hour. The panels correspond to \hbeta, \hgamma, \hdelta,
He~\textsc{i} $\lambda$4471 and He~\textsc{i} $\lambda$4713 from top
to bottom. The error bars are the same as those shown in Figure
\ref{fg:f5} but are omitted for clarity.\label{fg:f6}}

\figcaption[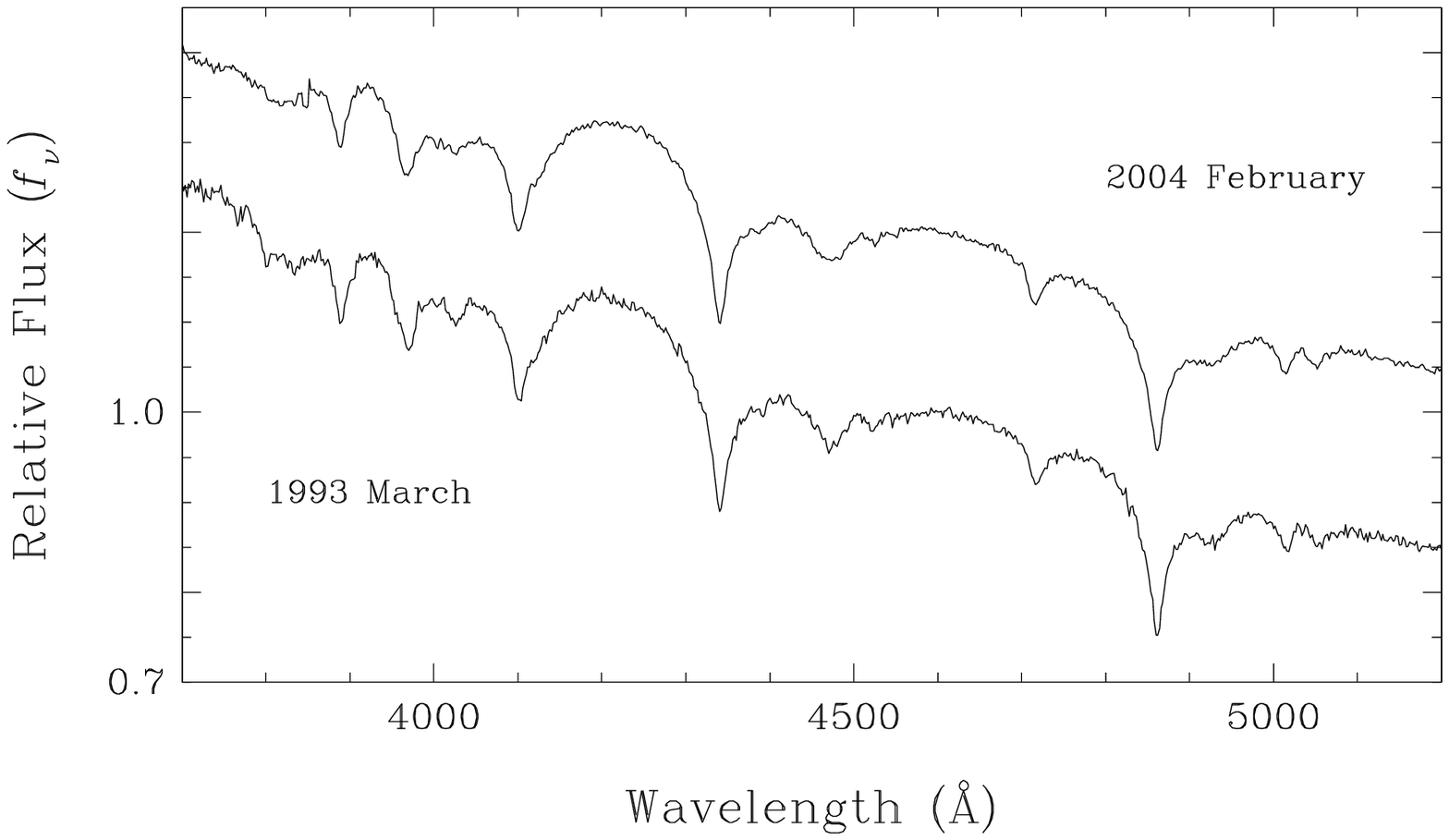]{Comparison of our template spectrum of GD 323 
(average of 27 spectra) with that generated from the data of
\citet{kls94} for the night of 1993 March 31 (average of 6
spectra). The spectra are normalized to unity at 4600~\AA\ and offset
from each other by a factor of 0.2.  The instrumental setup was
similar between the two sets of observations, but the reduction
procedure was different (see text).\label{fg:f7}}

\figcaption[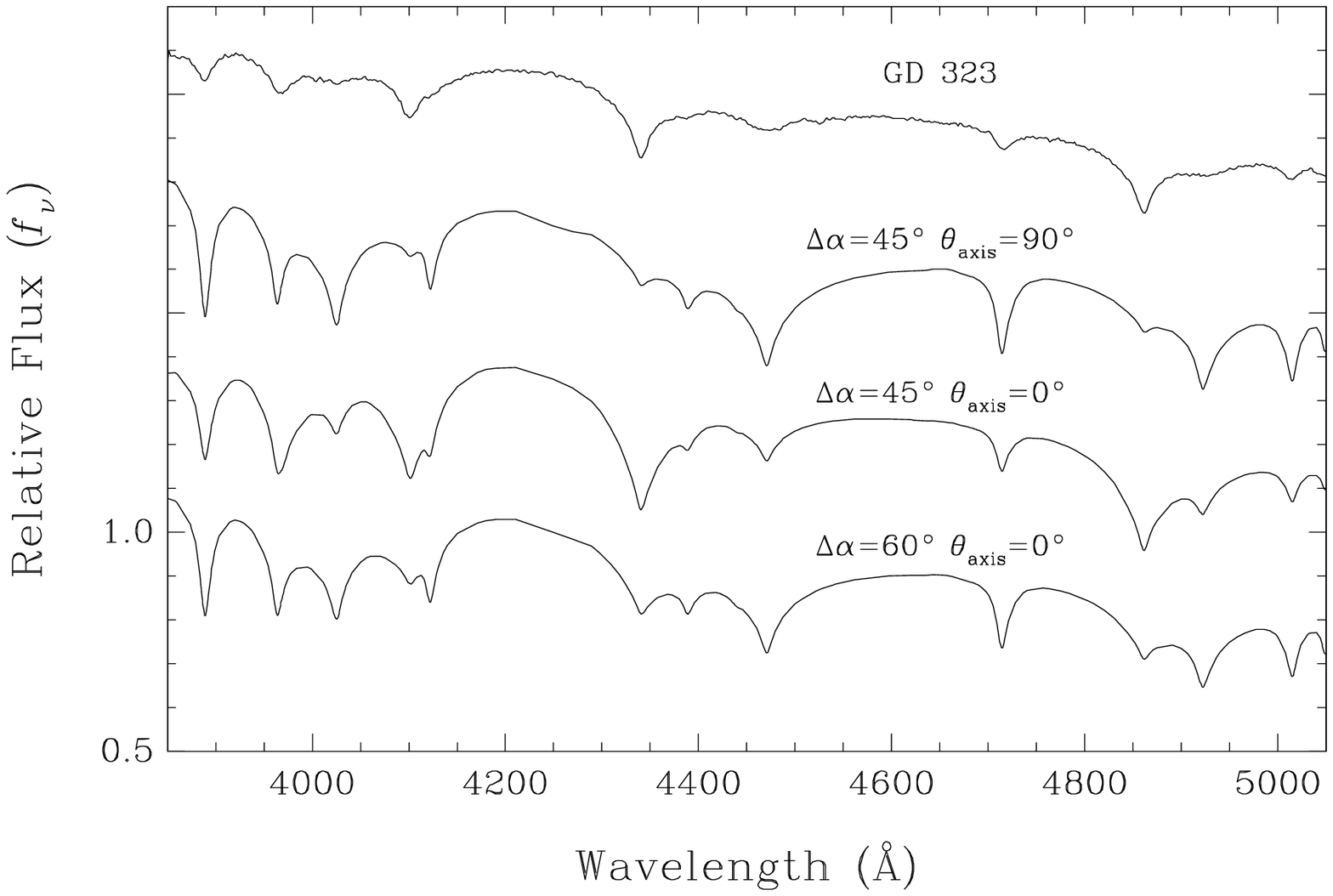]{Sample optical spectra of models with a pure helium 
equatorial band and pure hydrogen polar caps; the top spectrum is that
of GD 323 (average of 27 observations). The spectra are normalized to
unity at 4250~\AA\ and offset from each other by a factor of 0.35.
The half-width of the pure helium equatorial band is given by
$\Delta\alpha$, while $\theta_{\rm axis}$ represents the angle between
the stellar symmetry axis and the line of sight.\label{fg:f8}}

\clearpage

\begin{figure}[p]
\plotone{f1.eps}
\begin{flushright}
Figure \ref{fg:f1}
\end{flushright}
\end{figure}

\begin{figure}[p]
\plotone{f2.eps}
\begin{flushright}
Figure \ref{fg:f2}
\end{flushright}
\end{figure}

\begin{figure}[p]
\plotone{f3.eps}
\begin{flushright}
Figure \ref{fg:f3}
\end{flushright}
\end{figure}

\begin{figure}[p]
\plotone{f4.eps}
\begin{flushright}
Figure \ref{fg:f4}
\end{flushright}
\end{figure}

\begin{figure}[p]
\plotone{f5.eps}
\begin{flushright}
Figure \ref{fg:f5}
\end{flushright}
\end{figure}

\begin{figure}[p]
\plotone{f6.eps}
\begin{flushright}
Figure \ref{fg:f6}
\end{flushright}
\end{figure}

\begin{figure}[p]
\plotone{f7.eps}
\begin{flushright}
Figure \ref{fg:f7}
\end{flushright}
\end{figure}

\begin{figure}[p]
\plotone{f8.eps}
\begin{flushright}
Figure \ref{fg:f8}
\end{flushright}
\end{figure}

\end{document}